\documentclass[aps,prl,reprint,superscriptaddress,twocolumn,longbibliography]{revtex4-1}

\usepackage{graphicx}  
\usepackage{soul}
\usepackage{color}

\begin{document}


\title{Self-organized criticality in single neuron excitability}


\author{Asaf Gal}
\email[Corresponding author: ]{asaf.gal@mail.huji.ac.il}

\affiliation{The Interdisciplinary Center for Neural Computation (ICNC), The Hebrew University, Jerusalem, Israel}
\affiliation{Network Biology Research Laboratories, Lorry Lokey Interdisciplinary Center for Life Sciences and Engineering, Technion, Haifa, Israel}
\author{Shimon Marom}\thanks{The authors thank Erez Braun, Dani Dagan and Yariv Kafri for insightful comments and discussions. The research leading to these results has received funding from the European Unions Seventh Framework Program FP7 under grant agreement 269459.}
\affiliation{Network Biology Research Laboratories, Lorry Lokey Interdisciplinary Center for Life Sciences and Engineering, Technion, Haifa, Israel}
\affiliation{Department of Physiology, Faculty of Medicine, Technion, Haifa, Israel}
\thanks{The authors thank Erez Braun, Dani Dagan and Yariv Kafri for insightful comments and discussions. The research leading to these results has received funding from the European Unions Seventh Framework Program FP7 under grant agreement 269459.}
\date{\today}

\begin{abstract}
We present experimental and theoretical arguments, at the single neuron level, suggesting that neuronal response fluctuations reflect a process that positions the neuron near a transition point that separates excitable and unexcitable phases.  This view is supported by the dynamical properties of the system as observed in experiments on isolated cultured cortical neurons, as well as by a theoretical mapping between the constructs of self organized criticality and membrane excitability biophysics.
\end{abstract}


\maketitle


\subsubsection*{Introduction}

Cellular excitability is a fundamental physiological process whereby voltage-dependent changes in exciting and restoring membrane ionic conductances lead to an \textit{action potential} (AP), a transient change in trans-membrane voltage. Hodgkin and Huxley \cite{HODGKIN:1952ri} formalized a generic biophysical mechanism underlying the ignition and propagation of action potentials. In this formalism, as well as in its later extensions, the flow of ions down their electrochemical gradients is modulated by the probability of ion channel proteins to reside in a conductive state. 
An extensive set of observations shows that the activity and response properties of neurons are highly variable, fluctuating over extended time scales in a complex manner (e.g. \cite{Lowen:1996zv,Teich:1997ta,Gal2010a}) that are not easily accounted for by either the original Hodgkin-Huxley formalism nor by its extensions i.e.~by adding more ionic currents, gates and channel states, see \cite{Soudry2012}).  
Several approaches have been suggested for explaining these fluctuations, largely focusing on the stochastic nature of underlying mechanisms \cite{Schneidman:1998rc,Lowen:1999ad,Soen:2000fk,Gilboa:2005ej}, non linearity and chaotic dynamics \cite{Korn2003,Marom:2009ov}, or network level effects \cite{VanVreeswijk1996}.  

This paper approaches variability and complexity in single neuron activity from a different viewpoint. 
Conventional analyses and models of excitability use dynamical system approaches \cite{Izhikevich2007}.  However, excitability is known to be an emergent property of coupled states of numerous interacting microscopic elements - ion channels, calling for statistical-mechanics description.  While clues do exist for the potential benefit of thinking on excitability in statistical-mechanics terms (for instance, voltage fluctuations near the spiking bifurcation point, in biophysical models of excitability, were shown to exhibit critical-like behavior \cite{Roa2007,Steyn-Ross2006}), such an approach has not yet been explicitly proposed.
Here we apply the framework of self-organized criticality (SOC) to the level of a single, individual neuron, treated as an ensemble of \emph{interacting ion channels}.  

Originally introduced into the study of neural systems as a framework for explaining distributions of durations and sizes of network-wide events of activity (`avalanches') in cultured neural networks \cite{Beggs2003}, SOC has since been applied to \textit{in-vivo} recordings at the network and whole-brain levels \cite{Chialvo2010}. Acknowledging the controversies that surround its experimental foundations \cite{Dehghani2012,Bedard2006}, SOC provides an attractive theoretical framework for explaining the emergence of complexity in neural dynamics, establishing intriguing links to a long tradition of statistical-mechanics  treatment of neural networks \cite{Sompolinsky1988}. However, all of these studies treated the neural network as an ensemble of interacting \emph{neurons}, markedly different from the study presented here, which consider the neuron as  an isolated physical system. The relevance of our ideas to brain dynamics remains to be determined.    

\begin{figure}[t!]
\includegraphics[width=\columnwidth]{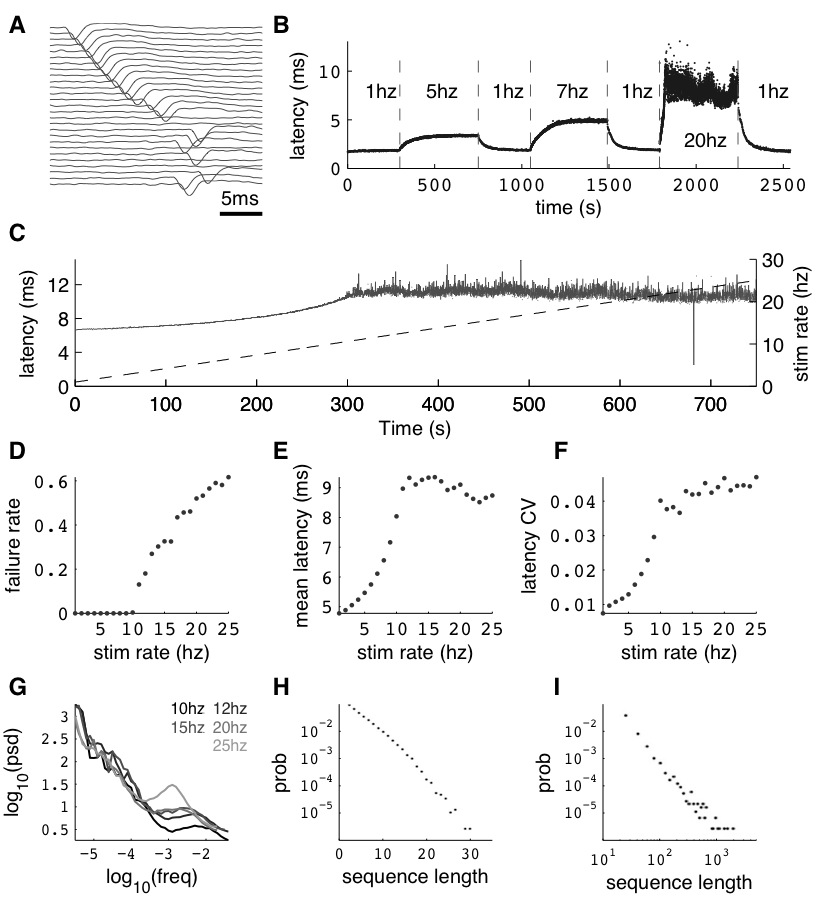}
\caption{Experimental observation of excitability dynamics.
(A) The response of a single isolated neuron to sequences of pulse stimuli delivered at 20Hz. The responses are ordered top to bottom, every 20th response is shown for clarity. The delaying of the AP can be observed, as well as response failures when excitability is below threshold. 
(B) The AP latency plotted as a function of time in an experiment where the stimulation rate is changed. For low stimulation rate, the excitability stabilizes at a fixed, supra threshold value. For high stimulation rate (20Hz), excitability decreases below threshold, and the neuron responds intermittently.
(C) response latencies (solid line) in response to a stimulation sequence with slowly increasing stimulation rate (dashed line). 
(D) Failure (no spike) probability as a function of stimulation rate. A critical stimulation rate is clearly evident. 
(E) Mean response latency as a function of stimulation rate. The increase of the latency accelerates as the stimulation rate approaches the critical point. 
(F) The jitter (coefficient of variation) of the latency as a function of stimulation rate.
(G) Scale free fluctuations in the intermittent mode. Periodograms of the failure rate fluctuations, at 5 different stimulation rates above $r_0$.
(H) Length distribution of spike-response sequences, on a semi logarithmic plot, demonstrating an exponential behaviour. Example from one neuron stimulated at 20Hz for 24 hours.
(I) Length distribution of no-spike response sequences from the same neuron, on a double logarithmic plot, demonstrating a power-law-like behavior. 
\label{fig:exp}}
\end{figure}

\subsubsection*{Summary of experimental observations}
In a series of experiments, detailed in a previous publication \cite{Gal2010a}, the intrinsic dynamics of excitability was observed by monitoring the responses of single neurons to series of pulse stimulations. In brief, cortical neurons from newborn rats were cultured on multi-electrode arrays, allowing extracellular recording and stimulation for long, practically unlimited durations. The neurons were isolated from their network by means of pharmacological synaptic blockage to allow study of intrinsic excitability dynamics, with minimal interference of coupled cells. Neurons were stimulated with sequences of short, supra-threshold identical electric pulses. For each pulse, the binary response (AP produced or not) was registered, marking the neuron as residing in either \emph{excitable} or \emph{unexcitable} state. For each AP recorded, the \emph{latency} from stimulation to the AP was also registered.
The amplitude of the stimulating pulses was constant and set well above threshold, such that the neuron will respond in a 1:1 manner (i.e. every stimulation pulse produces an AP) under a low rate (1Hz) stimulation condition. 

When the stimulation rate $r$ is increased to values higher than 1Hz, two distinct response regimes can be identified: a \emph{stable} regime, in which each stimulation elicits an AP, and an \emph{intermittent} regime, in which the spiking is irregular. The response of a neuron following a change of stimulation rate is demonstrated in Figure \ref{fig:exp}A and \ref{fig:exp}B, as well as in \cite{Gal2010a}: When stimulation rate is abruptly increased to a higher value, the latency gradually becomes longer and stabilizes on a new value. For high enough stimulation rate (above a critical value $r_0$), the 1:1 response mode  breaks down and becomes intermittent.  All transitions are fully reversible.
The steady state properties of the two response regimes may be observed by slowly changing the stimulation rate.  As seen in the result of the `adiabatic' experiment (Figure \ref{fig:exp}C), the stable regime is characterized by 1:1 response (no failures), stable latency (low jitter) and monotonous dependency of latency on stimulation rate. In contrast, the intermittent regime is characterized by a failure rate that increases with stimulation rate, unstable latency (high jitter) and independence of the mean latency on the stimulation rate. The existence of a critical (or threshold) stimulation rate is reflected in measures of failure rate (Figure \ref{fig:exp}D), mean latency (Figure \ref{fig:exp}E) and latency coefficient of variation (Figure \ref{fig:exp}F). The exact value of $r_0$ varies considerably between neurons, but its existence is observed in practically all measured neurons (see details in \cite{Gal2010a}). 

Within the intermittent regime, the fluctuations of excitability (as defined by the excitable/unexcitable state sequence) are characterized by scale-free long-memory statistics. Its power spectral density (PSD) exhibits a power-law ($1/f^\beta$) tail at the low frequency domain. The characteristic exponent of this power-law does not depend on the stimulation rate, as long as the latter is kept above $r_0$  (Figure \ref{fig:exp}G). The typical exponent of the rate PSD is $\beta=1.26\pm0.21$ (mean $\pm$ SD, calculated over 16 neurons). 
Moreover, within the intermittent regime, the distributions of the lengths of consecutive response sequences (i.e. periods of time the neuron is fully excitable, responding to each stimulation pulse) and consecutive no-response sequences (i.e. periods of time the neuron is not responding), are qualitatively different (Figures \ref{fig:exp}H and \ref{fig:exp}I).  The consecutive response sequence length histogram is strictly exponential, having a characteristic duration, while the consecutive no-response sequence length histogram is wide, to the point of  scale-freeness. Likelihood ratio tests for power law distribution fit to the empirical histogram (containing 90,000 samples) yielded significantly more likelihood comparing with Exponential fit, log normal, stretched exponential, and linear combination of two exponential distributions (all with normalized log likelihood ratios of $R>10$, $p<0.001$, see \cite{Clauset2009}). This suggests that the fluctuations are dominated by widely distributed excursions into an unexcitable state.\\ 

\subsubsection*{Interpretation}
In what follows we suggest an interpretation to the origin of the above experimental results in terms of critical phenomena, accounting for the complex statistics of single neuron excitability over extended time scales. The concept of `excitability' is vaguely defined; it generally reflects the susceptibility of the cell to produce an action potential in response to input above a given amplitude. In that sense, a `non-excitable' cell is one that can not evoke an AP, regardless of the stimulation amplitude, while 'excitable' cell is characterized by a continuous measure that quantifies excitability. Such a measure can be for example the minimal stimulation amplitude required for evoking an AP (the threshold), or alternatively the latency of the evoked AP, a more easily observed measure which is tightly related to the threshold.

Excitability is a lumped product of the individual states of numerous interacting ion channels; the aggregated macroscopic availability of these ionic channels to move into the conductive state and participate in the generation of action potentials.  
In the short term, the total number of ionic channels that are available to participate in AP generation may safely be assumed to be a constant. In the original Hodgkin and Huxley formalism, aims at the scale of milliseconds, the latter assumption is translated to maximal conductance \emph{parameters} that set limits on the instantaneous dynamics of the membrane. However, when long-term effects are sought, the maximal conductance might (and indeed should) be treated as a macroscopic system variable governed by stochastic, activity dependent, transitions of ion channels into and out from long-lasting \textit{unavailable} states.  These transitions are globally and locally coupled via membrane voltage, ionic concentrations and cellular modulatory and homeostatic processes. Since unavailable channels cannot contribute to membrane electrical response, slow changes in maximal conductance will be reflected in the time-amplitude envelope of the generated action potential, as well as in the very ability to generate it.  The precise impacts of slow changes in maximal conductances on excitability depend upon the specific type of ionic channel involved (i.e., mediating exciting or restoring ionic flows). Figures \ref{fig:sim}A and \ref{fig:sim}B exemplify this point for the case of sodium maximal conductance.  Such viewed, excitability has the flavor of an order-parameter, a measurable macroscopic physical quantity that reflects an average over the individual states of elements in an ensemble. The complex irregularity of neuronal responses over extended time scales, observed in the experiments described above, is thus naturally interpreted as a reflection of residency of the system near a phase transition between excitable and unexcitable phases, giving rise to the observed power-law statistics. 

However, given the above interpretation of fluctuations in excitability as reflecting critical phenomena, one would expect to observe the critical characteristics within a limited range of the experimental control parameter (i.e. stimulation rate); higher values of stimulation rate should shut-down excitability altogether.  This is not the case. For example, panels \ref{fig:exp}E and G show that response latency as well as the characteristic exponent of the power spectral density, are insensitive to the stimulation rate.  The reason for this apparent inconsistency is that stimulation rate does not directly impact on the dynamics of the underlying ionic channels. Rather, the relevant control parameter is in fact the \emph{activity} rate, itself a dynamic variable of the system.  This suggests a form of \emph{self-organization}. 

The concept of Self-Organized Criticality \cite{Bak1987} designates a cluster of physical phenomena characterizing systems that reside near a phase transition. What makes SOC unique is the fact that residing near a phase transition is not the result of a fine-tuned control parameter; rather, in SOC the system positions itself near a phase transition as a natural consequence of the underlying internal dynamic process that pushes towards the critical value. Such systems exhibit many complex statistical and dynamical features that characterize behavior near a phase transition, without these features being sensitive to system parameters. Dickman and his colleagues \cite{Dickman1998,Dickman2000} formalized a scheme for generating SOC from a conventional system exhibiting a phase transition. They have shown that many of the canonical models of SOC, including sandpile and forest fire models, can be understood as systems exhibiting absorbing state phase transition. Systems such as the contact process or activated random walk may reside about the phase transition, if amended with a carefully designed feedback: dissipating energy whenever the system is supercritical (i.e. permanently active without settling into an absorbing state), and driving the system whenever it is subcritical (i.e. when and only when it settles into the absorbing state).  

This picture naturally maps into excitability dynamics, where neural activity serves as a temperature-like parameter, and the single AP serving as a drive (quantal influx of energy, or small increase in temperature).  
In the absence of activity, the neuron reaches an excitable state (the `absorbing state' in such a mapping), while increased activity reduces excitability, and (when high enough) pushes the membrane into the unexcitable state. Residency in the unexcitable state decreases neural activity, leading to restoration of excitability. As a result, the neuron is kept around a barely-excitable state, exhibiting characteristics of SOC. 
Of course, not all classes of neurons follow this simplistic process, but the general idea holds:  activity pushes excitability towards a threshold state, while at the longer time scale regulatory feedback pulls the system back.\\ 

\subsubsection*{Model}
The above interpretation of excitability in SOC terms, may also be theoretically supported, within certain limits, by considering the underlying biophysical machinery. The state of the membrane is a function of the individual states of a large population of interacting ion channel proteins. A single ion channel can undergo transformations between uniquely defined conformations, conventionally modeled as states in a Markov chain. The faster transition dynamics between states is the foundation of the Hodgkin Huxley model, which describes the excitation event itself - the action potential. But, as explained above, for the purpose of modeling the dynamics of excitability, rather than the generative dynamics of the action potential itself, it is useful to group these conformations into two sets \cite{Toib:1998uk,Gilboa:2005ej,Marom:2009ov,Marom2010}: the \textit{available}, in which channels can participate in generation of action potentials, and the \textit{unavailable}, in which channels are deeply inactivated and are ``out of the game" of action potential generation. The microscopic details of the single channel dynamics in this state space, and definitely the collective dynamics of the interacting ensemble, are complex \cite{Liebovitch1987,Millhauser1988a,Millhauser1988} and no satisfactory embracing model exists to date. 
\begin{figure}[t!]
\includegraphics[width=\columnwidth]{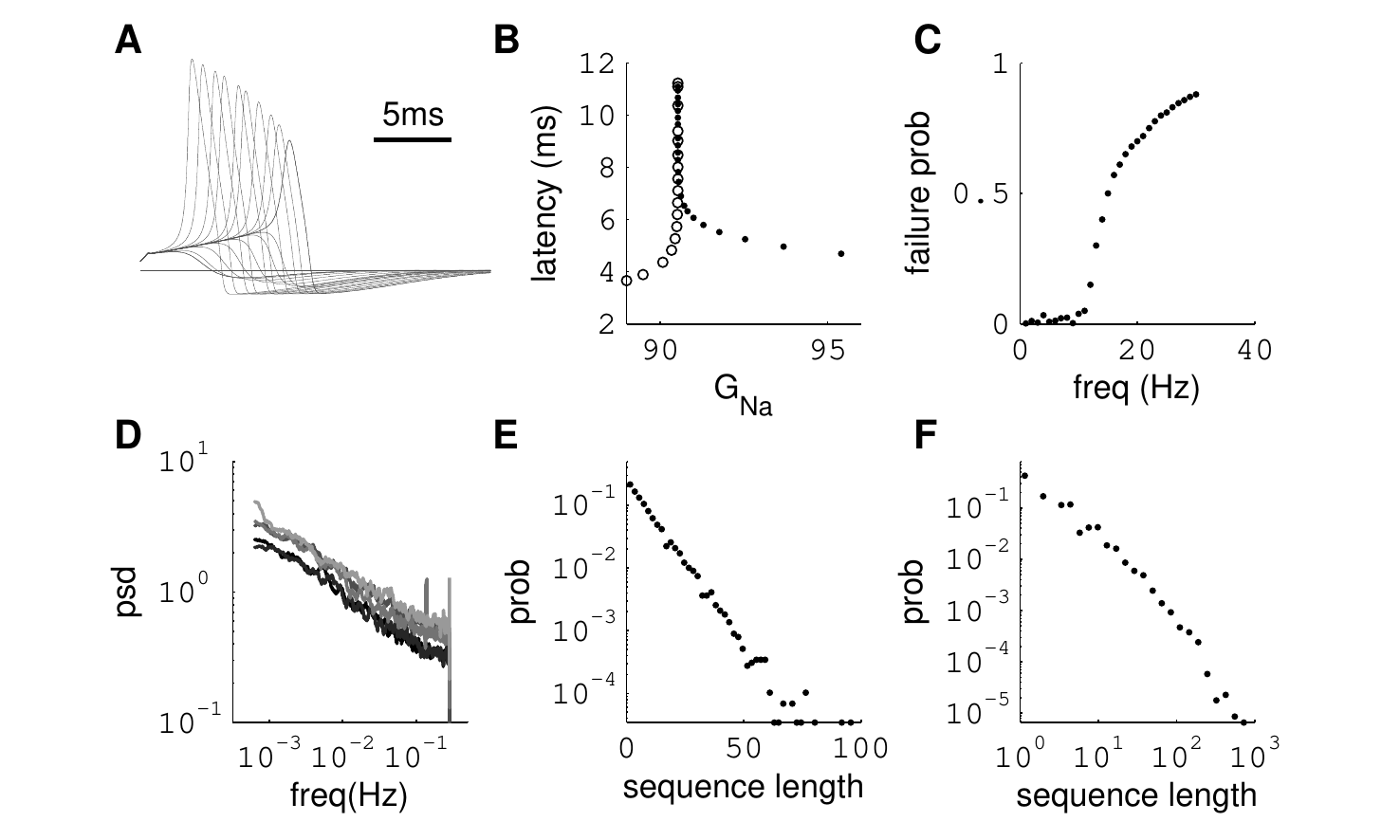}
\caption{
(A) The effect of modulating $G_{Na}$ in Hodgkin-Huxley model under short pulse stimulation. As it decreases, the AP is delayed. Below a certain threshold, no AP is produced. 
(B) AP latency in panel A as a function of $G_{Na}$, demonstrating the existence of a sharp threshold. Propagating APs are marked with filled circles, non-propagating responses are marked with empty circles.  
(C) Simulation results of the contact process model (equation \ref{eq2}). Dependence of the spike failure probability on the stimulation rate, analogous to Figure \ref{fig:exp}D. 
(D) Power spectral densities of the response fluctuations at different frequencies above the $r_0$, $1/f$-type behavior. 
(E) Length distribution of spike-response sequences, on a semi logarithmic plot, demonstrating an exponential behavior, analogous to Figure \ref{fig:exp}H.
(F) Length distribution of no-spike response sequences from the same neuron, on a double logarithmic plot, demonstrating a power-law-like behavior, analogous to Figure \ref{fig:exp}I.
\label{fig:sim}}
\end{figure} 
However, it has been suggested recently \cite{Marom:2009ov,Marom2010} that the transition dynamics between the available and unavailable states may be expressed in terms of an ``adaptive rate'', Logistic-like model of the general form:
 \begin{equation}
\dot{x} = -f(\gamma) x + g(x)(1-x),
\label{eq1}
\end{equation}
where $f$ is a function of the neural activity measure $\gamma$, and $g(x)$ is a monotonically increasing  function of the system state $x$, which represents the gross availability of ion channels.  In this modeling approach, complex transition statistics are a result of ensemble-level interaction, rather than of the internal structure of the single channel state space.

Following the lead of the above adaptive rate approach, one can consider, for instance, a model in which $x$ represents the availability of a restoring  (e.g.~potassium) conductance~\footnote{For example, the calcium dependent potassium SK channel \cite{Adelman2012} is an excitability inhibitor, having a calcium mediated positive interaction that gives rise to a form similar to equation~\ref{eq1}}. The state of the single channel is represented by a binary variable $\sigma_i$; $\sigma_i=0$ is the unavailable state and $\sigma_i =1$ is the available state. Unavailable channels are recruited with a rate of $x$, while available channels are lost with a rate of $2-\gamma$.  This picture gives rise to a dynamical mean field like equation: 
\begin{equation}
\dot{x} = (\gamma-1) x - x^2.
\label{eq2}
\end{equation}
The model is a variant of a globally coupled contact process, a well-studied system exhibiting an absorbing state phase transition \cite{Harris1974}. Here, $x=0$ is the absorbing state, representing the excitable state of the system. In the artificial case of $\gamma$ as an externally modified control parameter, for $\gamma<1$ (low activity) the system will always settle into this state, and the neuron will sustain this level of activity. For $\gamma>1$, the system will settle on $x^*=\gamma-1$, an unexcitable state, and the neuron will not be able to sustain activity.  
Feedback is introduced into the system by specifying the state dependency of $\gamma$: An AP is fired if and only if the system is excitable (i.e. in the absence of restoring conductance, $x$=0), giving rise to a small increase in $\gamma$. When $x>0$, the system is unexcitable, APs are not fired, and $\gamma$ is slowly decreased. This is an exact implementation of the scheme proposed in \cite{Dickman1998,Dickman2000}: an absorbing state system, where the control parameter (activity, $\gamma$) is modified by a feedback from the order parameter (excitability,  a function of $x$). 

As always with SOC, the distinction between order and control parameters becomes clear  only when the conservative, open-loop version of the model is considered.
Note that the natural dependency of the driving event (the AP) on the system state in our neural context, resolves a subtlety involved in SOC dynamics: the system must be driven slowly enough to allow the absorbing state to be reached, before a new quantum of energy is invested. In most models, this condition is met by taking driving rate to be infinitesimally small. 

Numerical simulation of the model (equation \ref{eq2}, together with the closed loop dynamics of $\gamma$, see methods section) qualitatively reproduces the power-law statistics observed in the experiment, including the existence of a critical stimulation rate $r_0$ (Figure \ref{fig:sim}C), the $1/f$ behavior for $r>r_0$, with exponent independent on $r$ (Figure \ref{fig:sim}D), and the distributions of sequence durations (Figures \ref{fig:sim}E and \ref{fig:sim}F).
The model has three relevant parameters: the integration timescale $\tau_{\gamma}$ of neural activity, the quantum of activity ($d\gamma$) added following each AP, and the stimulation rate $r$. The critical stimulation rate $r_0$, is adjustable by changing the first two parameters, and the SOC behavior is observed for any $r>r_0$, conditioned that $\tau_{\gamma} \gg 1/r_0$. While the model does capture key  observed properties, others are not accounted for. The latency transient dynamics when switching between stimulation rates (Figure \ref{fig:exp}B) and the multitude of stable latency values for $r<r_{0}$ (Figure \ref{fig:exp}C), suggest that a model with a single excitable state is not sufficient.  Sandpile models (and more generally activated random walk models, see \cite{Dickman1998,Dickman2000}), do exhibit such multiplicity, arising due to a continuum of stable subcritical values of pile height (or slope).  In this analogy, adding grains to the pile increases its height up to the critical point, where SOC is observed.  Another experimentally observed property that is not accounted for by the model, is the existence of \emph{pattern modes} in the intermittent response regime as described in \cite[Figure 10 of][]{Gal2010a}, implying strong temporal correlations between events of excitability and unexcitability.  These temporal correlations affect the exponent of the power-law spectral density, and might explain the difference between the exponent in the experiment and in the model simulation. Such correlations are not alien to SOC, and might arise in variant models \cite{Davidsen2002}.  

\subsubsection*{Concluding comments}

We have given several arguments, experimental and theoretical, in support of a connection between the framework of SOC and the dynamics underlying response fluctuations in single neurons. This interpretation succeeds in explaining critical-like fluctuations of neuronal responsiveness over extended timescale, which are not accounted for by other, more common, approaches \cite{Englitz2008,Soudry2012}. The key component that enables SOC in the ion channel ensemble is the existence of inter-channel interaction. While the interaction chosen here is \emph{global}, there are evidence that \emph{short range} cooperation  (as is more abundant in physical models of SOC) also exist between ion channels \cite{Naundorf2006}, and might be used to construct alternative models. 
Naturally, the simple model leading to equation \ref{eq2} is not unique, probably wrong in its microscopical details. Moreover, excitability is determined by more than one order parameter, and the interaction types are much more heterogeneous, controlled by an aggregate of such equations, representing the exciting and restoring forces, each pushing-pulling excitability to opposite directions.   

Nevertheless, while respecting the gap between theoretical models and biological reality, SOC seems to capture the core phenomenology of fluctuating neuronal excitability, and has a potential to enhance our understanding of physiological aspects of excitability dynamics.

\subsubsection*{Materials and Methods}

\paragraph*{Cultured neurons experiments.}
Experiments were performed on cultures of cortical neurons of newborn rats, as described in \cite{Marom:2002dg,Gal2010a}. Neurons were cultured on multielectrode arrays, allowing for extracellular recording of neuronal activity and extracellular electrical stimulation. As described in \cite{Gal2010a}, experiments where performed under complete blockage of synaptic transmission, to allow the study of intrinsic excitability dynamics in isolation from the effect of the activity of other neurons.  
Neurons were stimulated with sequences of short (400$\mu s$) pulse stimulations from one of the electrodes, with fixed inter stimulus interval. Following each pulse, the response of neurons was recorded (from a different electrode): whether a spike was fired or not, and the latency of the response from the stimulation pulse. While the effect of the stimulation is local to the part of the neuron near the stimulation electrode, the latency to the response reflects the conductance properties along the neuron from the stimulation electrode to the recording one, usually hundreds of microns away. Careful measures were applied  \cite{Gal2010a} to exclude experimental instabilities modulation the response of neurons over time. 

\paragraph*{Hodgkin-Huxley model simulation.}
A Hodgkin-Huxley model neuron (Figures \ref{fig:sim}A and \ref{fig:sim}B) was simulated using standard dynamic and rate equations \cite{DayanAbbott2001}. The neuron was stimulated with an injected rectangular current pulse (500$\mu s$ duration, $50\mu A$ amplitude), and the voltage dynamics was observed. Leak conductance $G_L$ was set to 0.3$mS$, the potassium conductance $G_K$ to 28$mS$ and the sodium conductance  $G_{Na}$ was changed in the range of 80-110$mS$. 

\paragraph*{Contact process simulation.}
Simulation was performed using an ensemble of 10,000 channels. The loop on neuronal activity $\gamma$ was closed as follows: for each AP fired, a single channel was inactivated, and $\gamma$ was increased by a value of $d\gamma=0.005$. Between APs, $\gamma$ decayed exponentially with a rate of $0.001$. Simulation length was 1 hour for each stimulation rate for Figure \ref{fig:sim}C, and 12 hours for each stimulation rate for Figures \ref{fig:sim}D-F. The full Matlab code of the simulation can be accessed in the authors website \footnote{here comes a link to the code}

\bibliography{Gal2012b}

\end{document}